\documentclass[pra,showpacs,superscriptaddress,amsmath,amssymb]{revtex4}
\usepackage{graphicx}

\begin{document}

\title{Efficient scheme for one-way quantum computing in thermal cavities}

\author{Wen-Xing Yang}
\affiliation{Department of Physics, Southeast University, Nanjing 210096, China} \affiliation{State Key Laboratory of Magnetic Resonance and Atomic and
Molecular Physics, Wuhan Institute of Physics and Mathematics, Chinese Academy of Sciences, Wuhan 430071, China}
\author{Zhe-Xuan Gong}  \email{gongzhexuan@gmail.com}
\affiliation{Department of Physics, Huazhong University of Science and Technology, Wuhan 430074, China}

\date{\today}

\begin{abstract}
We propose a practical scheme for one-way quantum computing based on efficient generation of 2D cluster state in thermal cavities. We achieve a
controlled-phase gate that is neither sensitive to cavity decay nor to thermal field by adding a strong classical field to the two-level atoms. We show that a
2D cluster state can be generated directly by making every two atoms collide in an array of cavities, with numerically calculated parameters and appropriate
operation sequence that can be easily achieved in practical Cavity QED experiments. Based on a generated cluster state in Box$^{(4)}$ configuration, we then
implement Grover's search algorithm for four database elements in a very simple way as an example of one-way quantum computing.
\end{abstract}


\maketitle

Over the past few years, the construction of a practical quantum computer has become a challenging goal for experimentalists. It is well known that the
building blocks of a general quantum computer are single-qubit rotations and two-qubit quantum gates \cite{1}. Recently, Briegel and Raussendorf \cite{2,3}
proposed a new idea for constructing quantum computer, known as one-way quantum computing, which shows that preparation of a particular entangled state, called
cluster state, accompanied with local single qubit measurements, is sufficient for simulating any arbitrary quantum logic operations. Cluster state as a
universal resource for general quantum computing has drawn extensive research interests \cite{4}-\cite{8}. Moreover, one way quantum computing by optical
elements based on four-qubit cluster states was recently demonstrated experimentally \cite{9}. It is hoped that experimental difficulties in performing complex
quantum gates may be overcome by one-way quantum computing based on the generation of cluster state.

A cluster state $|\psi\rangle_c$ can be visualized as a collection of qubits positioned at certain sites of a 2D lattice structure with lines connecting them,
which can be specified by the following set of eigenvalue equations:
\begin{equation}
\label{eq1} K^{(a)}|\psi\rangle_c = ( - 1)^{\kappa _a }|\psi\rangle_c
\end{equation}
with the correlation operators
\begin{equation}
\label{eq2} K^{(a)} = \sigma _x^{(a)} \mathop \otimes \limits_{b \in nghb(a)} \sigma _z^{(b)}
\end{equation}
where \textit{nghb(a)} is set of all the neighbors of any site $a$ of the lattice, and $\kappa _a \in \{0,1\}$. To generate an arbitrary cluster state, one can
first initialize each qubit in state $|+\rangle =(|0\rangle +|1\rangle)/\sqrt{2}$, where $|0\rangle$ and $|1\rangle$ are the computational basis, and then
perform controlled-phase operations between all neighboring qubits connected by the lines of the lattice.

Cavity QED system is known to be a qualified candidate for quantum information processing \cite{10}. However, up to now, one-way quantum computing based on
Cavity QED techniques has neither been proposed theoretically nor been carried out experimentally. The main difficulty lies in generating an arbitrary 2D
cluster state. Proposed schemes of generating cluster states using Cavity QED methods \cite{11}-\cite{14} are difficult for practical and scalable experiments
either due to the decoherence of the cavity field mode or due to the sensitivity of thermal field. Besides, most of the schemes are mainly focused on linear
cluster state prepared in one dimension, which are not suitable for use as substrate for quantum computation since one-way quantum computing based on 1D
cluster state can be efficiently simulated by classical computer, \cite{15,16} and most proposals for generating 2D cluster state are inefficient, as they
first need to generate several 1D cluster states and then collide them into a 2D configuration.

In the present work, we propose a practical scheme for one-way quantum computing based on efficient generation of 2D cluster state in thermal cavities.
Compared to Ref.\cite{11}-\cite{14}, our scheme is neither sensitive to cavity decay nor to thermal field, since the evolution of the atomic states is
independent of the cavity field mode, which is achieved by adding a strong classical field to our system. In addition, our implementation of controlled-phase
gate does not need any auxiliary state, \emph{i.e.} two-level atoms are used instead of three-level atoms, which further reduces experimental difficulties. On
the other hand, the 2D cluster state is generated in a direct and efficient manner in our scheme by appropriately choosing the initial velocity of each atoms
as well as the time delay between atom preparations and placing an array of cavities at certain locations in the path atoms passing through, so that every two
atoms can collide in a certain cavity and be subjected to entanglement generation as in Ref.\cite{17}. We give the generation of arbitrary 4-qubit cluster
state as an example with reasonable parameters and concrete operation sequence, and show that our scheme can perform one-way quantum computing process such as
Grover's search algorithm for four database elements in a simple and convenient way that is within the current experimental techniques.

Our generation of entanglement in cluster state is based on the interaction between two identical two-level atoms and a single-mode cavity field driven by a
classical field. In the rotating wave approximation, the Hamiltonian for such a system is given by (assuming $\hbar = 1)$ \cite{18}
\begin{equation}
\begin{array}{l}
\label{eq3} H = \sum\limits_{j = 1}^2 {\omega _0 \sigma _{z,j} } + \omega _a a^\dag a + \frac{1}{2}\sum\limits_{j = 1}^2 [g(a^\dag \sigma _j^ - + a\sigma _j^ +
) \\\\\quad\quad + \Omega (\sigma _j^ + e^{ - i\omega t} + \sigma _j^ - e^{i\omega t})]
\end{array}
\end{equation}
where $a$ and $a^\dag $ are the annihilation and creating operators for cavity mode, $\sigma _{z,j} = | e \rangle _j { }_j\langle e | - | g \rangle _j {
}_j\langle g |$, $\sigma _j^ + = | e \rangle _j { }_j\langle g |$, $\sigma _j^ - = | g \rangle _j { }_j\langle e |$, with $| e \rangle _j (| g \rangle _j )$
being the excited (ground) state of the $j$th atom. $\omega _0 $, $\omega _a $ and $\omega $ are the frequencies for atomic transition, cavity mode, and
classical field respectively. $g$ is the atom-cavity coupling strength and $\Omega $ is the Rabi frequency of the classical field. Assume that $\omega _0 =
\omega $. Then we can obtain the following interaction Hamiltonian in the interaction picture:
\begin{equation}
\label{eq4} H_i = \sum\limits_{j = 1}^2 {[\frac{\Omega }{2}(\sigma _j^ + + \sigma _j^ - ) + \frac{g}{2}(e^{ - i\delta t}a^\dag \sigma _j^ - + e^{i\delta
t}a\sigma _j^ + )]} ,
\end{equation}
with $\delta = \omega _0 - \omega _a $. For the new atomic basis $| \pm \rangle _j = (| g \rangle _j \pm | e \rangle _j ) / \sqrt 2 $, then we make a further
transformation with rotation with respect to the terms regarding $\Omega $ in Eq.(\ref{eq4}), and obtain

\begin{equation}
\begin{array}{l}
\label{eq5} H_I =\frac{g}{4}\sum\limits_{j = 1}^2 (| + \rangle _j { }_j\langle + | - | - \rangle _j { }_j\langle - | + | + \rangle _j { }_j\langle -
|e^{i\Omega t}
\\\\\quad\quad - |- \rangle _j { }_j\langle +
|e^{ - i\Omega t})e^{ - i\delta t}a^\dag + H.c.
\end{array}
\end{equation}

Free Hamiltonian $H_0 = \frac{\Omega }{2}\sum\limits_{j = 1}^2 {(| + \rangle _j { }_j\langle + | - | - \rangle _j { }_j\langle - |)} $ has been used
here for the transformation. Assuming that $\Omega \gg \delta ,g$, we can neglect the fast oscillating terms. Then obtain the effective interaction Hamiltonian,

\begin{equation}
\label{eq6} H_e = \frac{g}{2}(e^{ - i\delta t}a^\dag + e^{i\delta t}a)\sigma _x
\end{equation}
where $\sigma _x = \frac{1}{2}\sum\limits_{j = 1}^2 {(\sigma _j^ + + \sigma _j^ - )} $. The evolution operator for Hamiltonian (\ref{eq6}), which was first
proposed for trapped-ion system \cite{19}, can be written as
\begin{equation}
\label{eq7} U_e (t) = e^{ - iA(t)\sigma _x^2 }e^{ - iB(t)\sigma _x a}e^{ - iC(t)\sigma _x a^\dag }
\end{equation}

By solving the Schr\"{o}dinger equation $idU_e (t) / dt = H_i U_e (t)$, we can obtain $B(t) = g(e^{i\delta t} - 1) / 2i\delta $, $C(t) = - g(e^{ - i\delta t} - 1)
/ 2i\delta $, $A(t) = g^2[t + (e^{ - i\delta t} - 1) / i\delta ] / 4\delta $. Choosing $\delta t = 2\pi $, we have $B(t) = C(t) = 0$. Thus we get the evolution
operator of the system,

\begin{equation}
\label{eq8} U(t) = e^{ - iH_0 t}U_e (t) = e^{ - i\Omega t\sigma _x - i\lambda t\sigma _x^2 }
\end{equation}
where $\lambda = g^2 / 4\delta $. If we first apply single-qubit Hadamard gate on both atoms, and then set the interacting time $t$ and Rabi frequency $\Omega
$ appropriately so that $\lambda t = \pi / 2$ and $\Omega t = (2k + \frac{1}{2})\pi $ ($k$ is integer), followed by implementing again the Hadamard gate on
both atoms, then we obtain a controlled quantum phase gate with computational basis $|0\rangle,|1\rangle$ represented by $|e\rangle,|g\rangle$:
\begin{equation}
\label{eq9} \left \{ {\begin{array}{l}
 H^{\otimes 2}U(t)H^{\otimes 2}| g \rangle _1 | g \rangle _2 = - | g\rangle _1 | g \rangle _2 \\\\
 H^{\otimes 2}U(t)H^{\otimes 2}| g \rangle _1 | e \rangle _2 = | g\rangle _1 | e \rangle _2 \\\\
 H^{\otimes 2}U(t)H^{\otimes 2}| e \rangle _1 | g \rangle _2 = | e\rangle _1 | g \rangle _2 \\\\
 H^{\otimes 2}U(t)H^{\otimes 2}| e \rangle _1 | e \rangle _2 = | e\rangle _1 | e \rangle _2
\end{array}} \right.
\end{equation}

\begin{figure}
  \includegraphics[width=9cm]{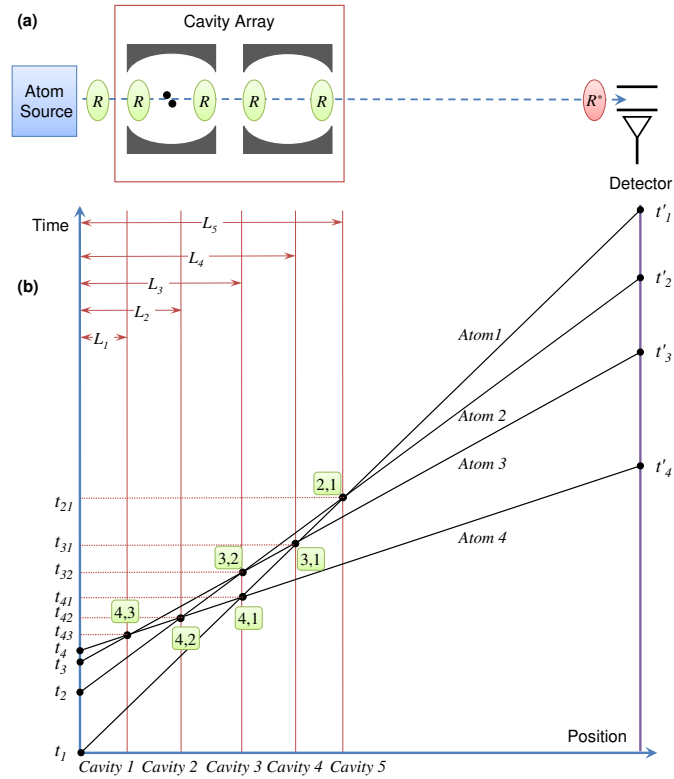}\\
  \caption{\textbf{(a)} Proposed experimental set-up for one-way quantum computing in an array of thermal cavities. R represents for Ramsey zone with $\pi/2$
resonant pulse added at the time of atom collisions. R$^*$ denotes for Ramsey zone with detuned $\pi/2$ pulse.\textbf{(b)} Space-time diagram for the sequence
of events. Every two atoms are made to cross the center of certain cavity simultaneously before finally reaching the detector.}\label{fig1}
\end{figure}

In order to generate an arbitrary two-dimensional cluster state using the controlled quantum phase gate given in Eq.(\ref{eq9}), we first assume that the
horizontal velocity $v_i$ and time $t_i$ of the $i^{th}$ atom emitting from the single-atom source has been pre-selected according to our need, and the
vertical position of each atom are slightly different. After the atoms initially in ground state move out horizontally from the source, a $\pi/2$ classical
resonant pulse R is added to each atom so that they are prepared in the state $(|0\rangle +|1\rangle)/\sqrt{2}$. Our next aim is to let every two atoms collide
in a certain cavity and so that they may undergo the evolution in Eq.(\ref{eq8}), and an arbitrary 2D cluster state can thus be formed. For N atoms forming a
N-qubit cluster state, calculations show that this requires us to place $(2N-3)$ cavities in a array as depicted in Figure \ref{fig1}(a). Suppose the $k^{th}$
cavity center is at a distance $L_k$ from the single-atom source, the following equations should be satisfied in order to achieve collisions between every two
atoms.
\begin{equation}\label{eq10}
L_{i+j-2}(\frac{1}{v_j}-\frac{1}{v_i})=t_i-t_j
\end{equation}
where $i=1,2,...,N-1;j=i+1,i+2,...,N$ and we set $t_1=0$ for simplicity. Eq.(\ref{eq10}) contains $N(N-1)/2$ nonlinear equations and $4(N-1)$ variables.
Calculations show that it has a group of solutions for $N\le6$ and may not give any solution for $N\ge7$. For pedagogical reasons, we are aimed at the
relatively simple but important case of $N=4$, where one can get many reasonable solutions that satisfy Eq.(\ref{eq10}). Figure \ref{fig1}(b) shows a possible
case qualitatively. The analytical solution we obtained is rather complicated, but our numerical calculation gives many quantitative solutions that are
appropriate in practical experiments. Table \ref{tab1} shows one example.
\begin{table}[here]
\caption{One numerical solution for Eq.(\ref{eq10}) in the case of $N=4$. Part of the quantities are set as constants and others are variables being solved.}\label{tab1}
\begin{tabular}{|c|c|c|c c|}
  \hline
  $v_1$ & $v_2$ & $v_3$ & $v_4$ & \\
  100m/s & 122m/s & 146m/s & 250m/s & \\ \hline
  $t_1$ & $t_2$ & $t_3$ & $t_4$ & \\
  0ms & 0.359ms & 0.471ms & 0.500ms & \\ \hline
  $L_1$ & $L_2$ & $L_3$ & $L_4$ & $L_5$ \\
  1.00cm & 3.35cm & 8.33cm & 15.00cm & 20.00cm \\
  \hline
\end{tabular}
\end{table}

Since we only want the atom pairs representing neighboring qubits in the cluster state to undergo the controlled-phase operation in Eq.(\ref{eq8}), in all the
other case, the cavity is set far off resonance by a large electric field applied across the cavity mirrors. This field stark-shifts the atomic levels far off
resonance, so that the atom-cavity interaction is then negligible \cite{20}. The interaction time in Eq.(\ref{eq8}) is also controlled strictly in the same way
together with control of the length of the strong classical pulse added. In fact, small difference of the interaction time between two atoms will not cause
measurable errors. \cite{18,21}

\begin{figure}
  \includegraphics[width=9cm]{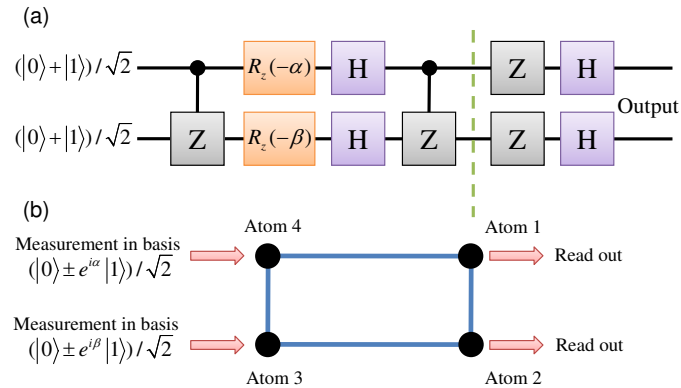}
  \caption{\textbf{(a)} Circuit model of Grover's search algorithm for four database elements. The two encoded qubits pass through a series of unitary quantum
logic gates before being measured in computational basis as output of the search algorithm. $\alpha,\beta$ are determined by the "Oracle", and setting
$\alpha\beta$ to $\pi\pi$,$\pi0$,$0\pi$,$00$ corresponds to the marked element encoded as $00$,$01$,$10$,$11$. \textbf{(b)} Cluster state quantum computing
model for the same algorithm. The physical qubits carried by Atom 4,3 are measured in any order, changing the states of Atom 2, 1 to states of the encoded
qubits in the circuit model at the same stage.}\label{fig2}
\end{figure}

Now we'd like to show how to implement one-way computing based on the generation of cluster state in our scheme. Here we use the Grover's search algorithm for
four unsorted database elements as an example, the circuit model \cite{22} and one-way computing model \cite{9} of which have been shown in Figure \ref{fig2}.
The cluster state we need to generate is in Box$^{(4)}$ configuration
($|\psi\rangle_c=|0\rangle_1|+\rangle_2|0\rangle_3|+\rangle_4+|0\rangle_1|-\rangle_2|1\rangle_3|-\rangle_4+|1\rangle_1|-\rangle_2|0\rangle_3|+\rangle_4+|1\rangle_1|+\rangle_2|1\rangle_3|-\rangle_4$),
which entails us to tune the cavity field to near resonance when atom pairs (4,3), (4,1), (3,2), (2,1) collide in Cavity 1, 3 and 5 at time
$t_{43},t_{41},t_{32},t_{21}$ as in Figure \ref{fig1}(b), while $\{v_1,v_2,v_3,v_4\},\{t_1,t_2,t_3,t_4\}$ and $\{L_1,L_2,L_3,L_4,L_5\}$ are
pre-set to values satisfying Eq.(\ref{eq10}), such as in Table \ref{tab1}.

After $t_{21}$, the cluster state has been generated and we are then to perform single-qubit measurements in the order of atom 4-3-2-1. Atom 4 and atom 3
should be measured in the basis $(|0\rangle \pm e^{i\alpha}|1\rangle)/\sqrt{2}$ and $(|0\rangle \pm e^{i\beta}|1\rangle)/\sqrt{2}$, which can be achieved
simply by adding $\pi/2$ Ramsey pulse with frequency $\omega_r$ satisfying $(\omega_r-\omega_0)T=\alpha(\beta)$ respectively, ($\omega_0$ is the atomic
transition frequency and $T$ is the coherent interaction time between the pulse and atom) and then measuring the atom in basis $\{|0\rangle,|1\rangle\}$
through Field Ionization Detector. Atom 2 and Atom 1 carry the read-out qubits in the cluster state model and should go through a $\sigma_z$ and a Hadamard
gate, which can be realized by adding the same $\pi/2$ Ramsey pulse with frequency $\omega_r$ satisfying $(\omega_r-\omega_0)T=\pi$, before final measurements
by the Detector in computational basis. If the measurement results of the four flying atom at time sequence $t'_4,t'_4,t'_2,t'_1$ are denoted by
$r_4,r_3,r_2,r_1$, then the one-way quantum computing process succeeds in giving us the search result encoded as $\{r_1\oplus r_3,r_2\oplus r_4\}$ if $r_3$ and
$r_4$ are not both zero \cite{9}.

\begin{figure}
  \includegraphics[width=6cm]{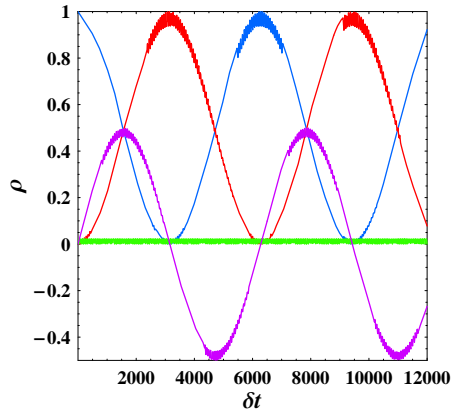}
  \caption{Coherent evolution of two atoms in a thermal cavity according to Eq.(\ref{eq8}), with the parameters $\delta=g$, $\Omega=5\delta$ and $n_{th}=1$.
The two atoms are assumed to be initially in the ground states $|g\rangle_1|g\rangle_2$. The red, blue, green and purple curvers represent for $\rho_{gg,gg}$,
$\rho_{ee,ee}$, Re($\rho_{gg,ee}$) and Im($\rho_{ee,ee}$) respectively. The small deviation from ideal situation on the curves are small oscillations with a
high frequency.}\label{fig3}
\end{figure}

We now discuss further the experimental feasibility of our one-way quantum computing scheme. From Eq.(\ref{eq8}), we note that the photon-number-dependent
parts in the evolution operator are canceled with the strong resonant classical field added, thus our scheme can be done in thermal cavities and is also
insensitive to cavity decay, greatly reducing experimental difficulties. The robustness of our scheme to thermal field is illustrated in Figure \ref{fig3}, where
we simulate the time evolution curves of the two atoms in the cavity by solving an appropriate master equation and allowing for heating in form of quantum
jumps described by jump operators $\sqrt {\Gamma n_{th} } a$ and $\sqrt {\Gamma (n_{th} + 1)} a^\dag $ ($\Gamma $ and $n_{th} $ are the typical heating rate
and the mean excitation of the photon). The Figure clearly shows that we have a coherent evolution of the atomic state which is not entangled with the cavity.
Besides, in obtaining Eq.(\ref{eq8}), there is no requirement that the atom-cavity detuning should be much larger than the atom-cavity coupling strength.
Operation time can be thus shortened, which is also important in view of decoherence.

Moreover, the velocities of the atoms can be selected by Doppler-selective optical pumping techniques, with a precision of $\pm 2m/s$, and the timing of each
atom preparation can be controlled to a precision of $2\mu s$ \cite{10}. Besides, the Cavity QED experimental apparatus are typically 20cm in length \cite{10}.
These validates our choice of $v_i$, $t_i$ and $L_i$. It is also worth mentioning that since the waist of the cavity field is at most a few mm, which is much
smaller than the distance between the cavities, it is unlikely that our space-time arrangement of atoms will cause more than two atoms simultaneously crossing
the center of the cavity field.

The radiative time for Rydberg atoms with principal quantum numbers around 50 is $T_r = 3\times 10^{ - 2}s$, and the coupling constant is $g = 2\pi \times
25$kHz \cite{10}. The corresponding photon storage time in a cavity can reach $T_c = 1$ms \cite{10}. In the present scheme, the virtually excited cavities have
only a small probability, about 1{\%}, of being excited during the passage of the atom pairs through them. Thus the efficient decay time of the cavity
$T_{ceff} \sim 0.1$s. Choosing $\delta = g$, direct calculation shows that the interaction time is on the order of $10^{ - 5}$s, which is much shorter than the
cavity decay time $T_{ceff} $. The time needed for the whole one-way quantum computing process can be controlled within a few ms according to the choice of
$t_i$, apparently smaller than the the radiative time of the Rydberg atoms $T_r$, rendering our scheme insensitive to the decoherence of atoms.

Errors in our scheme will mainly come from the process of entanglement generation in cavities and the single-qubit measurement process after generating the
cluster state. In the former case, the errors can be induced by Start shift on the states $| + \rangle _j $ and $| - \rangle _j $ as we've discarded the
fast-oscillating terms in obtaining Eq.(\ref{eq6}), and by pulse imperfections and initial cavity Fock state. Ref.\cite{23} shows that these errors will only
have slight influence on the fidelity of the entangled state obtained. In the latter case, we note that the measurement efficiency of ionization detectors
can be more than 80\% with current techniques \cite{24}. However, we're still expecting higher measurement efficiency in Cavity QED experiments.

For one-way computing of quantum algorithms that need no more than 6 physical qubits, our scheme does not need much change. However, for a large number of
qubits, Eq.(\ref{eq10}) cannot be satisfied by simply setting $\{v_i\}$, $\{t_i\}$, $\{L_i\}$ before the computing process. One of the possible ways to solve
this problem is to change the velocity of the atoms in the midway. Although Rydberg atoms are neutral and usually not easy to accelearte/deccelerate, we note
that recent progress on Cavity QED experiment with optically transported atoms can manipulate the motion of atoms by resorting the optical dipole force
\cite{25,26}. Such technique can provide a moderate acceleration for optically trapped atoms and deliver them deterministically, which might make our scheme
still possible to work for relatively large scale of one-way quantum computing process.

The authors would like to thank Prof. Ying Wu for many enlightening discussions and helpful suggestions. This work was partially supported by National Fundamental
Research Program of China 2005CB724508 and National Natural Science Foundation of China under Grant Nos. 60478029, 90503010, 10634060 and 10575040.

\end{document}